\documentclass[sn-mathphys-num]{sn-jnl}% sn-mathphys-num: Math and Physical Sciences Numbered Reference Style 

\usepackage{graphicx}%
\usepackage{multirow}%
\usepackage{amsmath,amssymb,amsfonts}%
\usepackage{amsthm}%
\usepackage{mathrsfs}%
\usepackage[title]{appendix}%
\usepackage{xcolor}%
\usepackage{textcomp}%
\usepackage{manyfoot}%
\usepackage{booktabs}%
\usepackage{algorithm}%
\usepackage{algorithmicx}%
\usepackage{algpseudocode}%
\usepackage{listings}%
\begin{document}

\title[Article Title]{Classical theory of nucleation applied to condensation of a Lennard-Jones fluid}

%%=============================================================%%
%% GivenName	-> \fnm{Joergen W.}
%% Particle	-> \spfx{van der} -> surname prefix
%% FamilyName	-> \sur{Ploeg}
%% Suffix	-> \sfx{IV}
%% \author*[1,2]{\fnm{Joergen W.} \spfx{van der} \sur{Ploeg} 
%%  \sfx{IV}}\email{iauthor@gmail.com}
%%=============================================================%%

\author*[1]{\fnm{Yijian} \sur{Wu}}\email{yijian.wu@polytechnique.edu}

\author[1]{\fnm{Thomas} \sur{Philippe}}\email{thomas.philippe@polytechnique.edu}

\affil*[1]{\orgdiv{Laboratoire de Physique de la Matière Condensée}, \orgname{CNRS, Ecole polytechnique, Institut Polytechnique de Paris}, \orgaddress{\city{Palaiseau}, \postcode{91120}, \country{France}}}

\abstract{The classical nucleation theory (CNT) and its modified versions provide a convenient framework for describing the nucleation process under the capillary approximation. However, these models often predict nucleation rates that depart significantly from simulation results, even for a simple Lennard-Jones fluid. This large discrepancy is likely due to the inaccurate estimation of the driving force for nucleation, which most traditional models estimate within the ideal solution approximation. 
In this study, we address this issue by directly calculating the driving force for nucleation using equations of state (EOS) and integrating this approach into the calculation of nucleation rates within the framework of CNT and its modified model. We apply this method to examine the condensation of a Lennard-Jones fluid and compare the resulting nucleation rates with molecular dynamics (MD) simulation data. 
Our results demonstrate that at relatively low supersaturation, where the capillary approximation is reasonable, our thermodynamic models exhibit  excellent agreement with MD results, significantly outperforming traditional models. 
At moderate and high supersaturation, our approach continues to show a reasonable agreement with MD results. Furthermore, when comparing the results obtained by using different EOS, we find that more precise EOS generally yield better agreement with MD data.}


%%\pacs[JEL Classification]{D8, H51}

%%\pacs[MSC Classification]{35A01, 65L10, 65L12, 65L20, 65L70}

\maketitle

\section{Introduction}\label{sec:Introduction}

Nucleation is a fundamental process with people's great interest in both basic science and material science. Despite its importance, our understanding of nucleation remains incomplete. In supersaturated vapor systems for instance, liquid clusters spontaneously form from monomers due to thermodynamic fluctuations. These clusters may either grow into large structures, or shrink back into smaller clusters or individual monomers. This process is known as condensation. The classical nucleation theory (CNT) \cite{kalikmanov_nucleation_2013} describes the nucleation process and calculates nucleation rates. Its simplicity and ease of application have made it widely used across various fields.

However, CNT has been increasingly challenged by experimental findings and numerical simulations. For instance, the steady-state nucleation rates predicted by CNT can differ from experimental results of argon by as much as 11 to 26 orders of magnitude \cite{iland_argon_2007,sinha_argon_2010}. Additionally, molecular dynamics (MD) and Monte Carlo (MC) simulations have shown that the nucleation rates obtained from numerical simulations deviate significantly from those predicted by CNT \cite{diemand_large_2013,yoo_monte_2001}. 
Although some modifications to CNT have been proposed \cite{ford_nucleation_1997,laaksonen_revised_1994}, they have not resolved these discrepancies. 

A critical component of CNT and its modified models is the work of formation of the critical cluster, i.e., the nucleation barrier. This quantity is crucial because it appears exponentially in the nucleation rate equation, making the nucleation rate highly sensitive to the accuracy of the critical work of formation. 

CNT relies on the capillary approximation, i.e. the embryo of the new phase is a sphere separated from the vapor phase by a sharp interface that gives rise to an energy penalty proportional to the interface area. The latter approximation is expected to be accurate at low supersaturation where critical clusters are relatively large. 
Diemand et al. \cite{yoo_monte_2001} conducted large-scale brute force MD simulations of homogeneous vapor-to-liquid nucleation in Lennard-Jones systems. Large domains allow for the examination of nucleation process at low supersaturation, down to 1.60. Their results showed that, even at relatively low supersaturation levels between 1.60 and 1.66, CNT and its modified models overestimate the nucleation rate by 2 to 4 orders of magnitude. This significant discrepancy is likely due to the inaccurate estimation of the driving force for nucleation, which is approximated by the logarithm of the supersaturation in these models.

In this paper, we incorporate a direct and more accurate method for calculating the driving force for nucleation into the nucleation rate formula within the framework of CNT and its modified model. 
In Section \ref{sec:Methods}, we review the calculation process of nucleation rate in traditional models and present the application of equation of states for determining the driving force for nucleation. In Section \ref{sec:Results}, we apply our method to Lennard-Jones system and compare our results with MD simulations results and the predictions from traditional models.

\section{Methods}\label{sec:Methods}

\subsection{Classical nucleation theory and its modified version}\label{subsec:CNT}
In a supersaturated vapor, liquid clusters may form due to thermodynamic fluctuations. In the capillary approximation of CNT, the thermodynamics of condensation can be easily described. The work required to form a cluster consists of contributions from the energetically favorable bulk free energy difference between the two phases, proportional to the volume of the cluster, and the unfavorable surface forming energy, proportional to the cluster's surface area. In CNT the clusters have the same number density as the bulk liquid, $\rho^\mathrm{l}$. Their sizes are characterized by the number of monomers that they contain, $n$, which serves as the sole variable in the theory. 
For a spherical cluster of size $n$, its volume and surface are given by $V = 4 \mathrm{\pi} r_0^3 n/3$ and $A = 4 \mathrm{\pi} r_0^2 n^{2/3}$, respectively, where $r_0$ is the radius of monomer. 
The work of formation of such a cluster is then expressed as 
\begin{equation}
    \Delta\varOmega_\mathrm{CNT}(n) = -n \Delta \mu + 4 \mathrm{\pi} r_0^2 n^{2/3} \gamma .
\label{eq:WOF CNT original}
\end{equation}
Here, $\Delta\mu$ represents the difference in chemical potential between the vapor phase and the liquid phase, and $\gamma$ is the surface tension of the cluster, often approximated by the surface tension of the flat interface. Both $\Delta\mu$ and $\gamma$ are positive quantities.

The work required to form a monomer as given by Eq. (\ref{eq:WOF CNT original}), $\Delta\varOmega_\mathrm{CNT}(1)$, does not vanish. For this purpose, the modified classical nucleation theory (MCNT) \cite{ford_nucleation_1997} was introduced. MCNT redefines the work of formation as follows:
\begin{equation}
    \Delta\varOmega_\mathrm{MCNT}(n) = -(n-1) \Delta \mu + 4 \mathrm{\pi} r_0^2 (n^{2/3}-1) \gamma .
\label{eq:WOF_MCNT}
\end{equation}

A cluster reaches its critical size, $n_\mathrm{c}$, when its work of formation reaches the maximum. The critical size is given by 
\begin{equation}
    n^* = \dfrac{32 \mathrm{\pi} \gamma^3}{3(\rho^\mathrm{l})^2(\Delta\mu)^3}
\label{eq:nc}
\end{equation}
in the framework of both CNT and MCNT.
The corresponding work of formation is the so-called critical work of formation, or nucleation barrier, given by
\begin{equation}
    \Delta\varOmega_\mathrm{CNT}^* = \dfrac{16\mathrm{\pi}\gamma^3}{3(\rho^\mathrm{l}\Delta\mu)^2} 
    \quad \mathrm{and} \quad
    \Delta\varOmega_\mathrm{MCNT}^* =  \dfrac{16\mathrm{\pi}\gamma^3}{3(\rho^\mathrm{l}\Delta\mu)^2} + \Delta\mu - 4\mathbf{\pi}r_0^2\gamma.
\label{eq:CNT DeltaOmega_c}
\end{equation}

The steady-state nucleation rate, defined as the number of clusters of critical size produced per unit volume and per unit time at steady state, can be calculated using the well established expression \cite{yasuoka_molecular_1998} in both CNT and MCNT models, 
\begin{equation}
    I = \dfrac{\rho_0^2}{\rho^\mathrm{l}} \sqrt{\dfrac{2\gamma}{\mathrm{\pi}m}} \exp{\Big(-\dfrac{\Delta\varOmega^*}{k_\mathrm{B}T}\Big)} ,
\label{eq:I}
\end{equation}
where $\rho_0$ is the number density of the supersaturated vapor, $k_\mathrm{B}$ is the Boltzmann constant, $T$ is the system temperature, and $m$ is the mass of a monomer, $m = 4\mathrm{\pi}\rho^\mathrm{l}r_0^3/3$.

\subsection{Determination of driving force for nucleation} \label{subsec:driving force}
A key challenge in calculating the nucleation rates lies in determining the chemical potential difference, $\Delta\mu$, which serves as the driving force for nucleation. Traditional models often assume that the vapor phase behaves as an ideal gas and that the liquid phase is incompressible \cite{kalikmanov_nucleation_2013}, leading to the simplified expression:
\begin{equation}
    \Delta\mu = k_\mathrm{B}T\ln{S} ,
\label{eq:lnS}
\end{equation}
where $S$ is the supersaturation ratio, defined as the ratio of the partial pressure of monomers to the saturated vapor pressure. 
While these assumptions simplify drastically calculations, they may not be accurate for most real substances. 
The semi-phenomenological (SP) model \cite{laaksonen_revised_1994} was developed to address the nonideality of the vapor phase based on MCNT. In this model, the work of formation is expressed as
\begin{equation}
    \Delta\varOmega_\mathrm{SP}(n) = -(n-1) \Delta \mu + 4 \mathrm{\pi} r_0^2 (n^{2/3}-1) \gamma + k_\mathrm{B}T\xi (n^{1/3}-1),
\label{eq:WOF_SP}
\end{equation}
where the additional parameter $\xi$ is determined from $\gamma$, $r_0$, $T$, the second virial coefficient, and the saturation pressure of the monomer vapor \cite{diemand_large_2013}.

Alternatively, a more precise approach to bypass the ideal gas assumption is to directly calculate the chemical potential difference using equations of state (EOS), which provide a comprehensive relationship between state variables, allowing for an accurate determination of the chemical potential. However, it is important to note that this direct calculation should not be applied to the SP model, as it would render the correction for vapor nonideality redundant.

\section{Results}\label{sec:Results}
In this study, we investigate systems governed by the well-known 12-6 Lennard-Jones potential, $v_\mathrm{LJ}(r) = 4\epsilon\big((\sigma/r)^{12} - (\sigma/r)^6\big)$, with $\epsilon$ and $\sigma$ the energy and length parameters of the potential, respectively. 
The parameters $\epsilon$, $\sigma$, Boltzmann constant $k_\mathrm{B}$, and monomer mass $m$ are used as the reducing parameters. 

The chemical potential difference, $\Delta\mu$, is calculated using EOS. Several different series of EOS exist for Lennard-Jones system \cite{stephan_review_2020}. In this study, we employ two of the most commonly-used series: EOS proposed by Kolafa and Nezbeda \cite{kolafa_lennard-jones_1994}, denoted as KN, and EOS proposed by Johnson, Zollweg, and Gubbins \cite{johnson_lennard-jones_1993}, denoted as JZG.

The surface tension $\gamma$ and the cluster density $\rho^\mathrm{l}$ are computed using the formulae proposed by Baidakov et al. \cite{baidakov_metastable_2007}. The monomer radius $r_0$ is determined by the relation $r_0 = (3m/(4\mathrm{\pi}\rho^\mathrm{l}))^{1/3}$. 

Diemand et al. \cite{diemand_large_2013} performed MD brute force simulations of Lennard-Jones system in large-scale domains, which allow them to obtain nucleation data at relatively low supersaturation. They directly measured the supersaturation and nucleation rates for each configuration, and also calculated nucleation rates by using the traditional MCNT and SP models based on their measured supersaturation. 

In this work, we calculate nucleation rates and critical cluster sizes for their configurations with temperatures $T \geq 0.6\epsilon/k_\mathrm{B}$ by applying EOS on CNT and MCNT, as given in Eq. (\ref{eq:I}) and (\ref{eq:CNT DeltaOmega_c}). We calculate nucleation rates by using traditional CNT model with the supersaturation obtained from simulations, and the work of formation for all models as well.
We limit our calculations to $T \geq 0.6\epsilon/k_\mathrm{B}$ because EOS may not accurately describe the thermodynamic properties at lower temperatures. 
It is important to note that, unlike the work of Tanaka et al. \cite{tanaka_free_2014}, our models do not involve any fitting parameters.  

\begin{figure}[hpbt]
\centering
\includegraphics[width=0.9\textwidth]{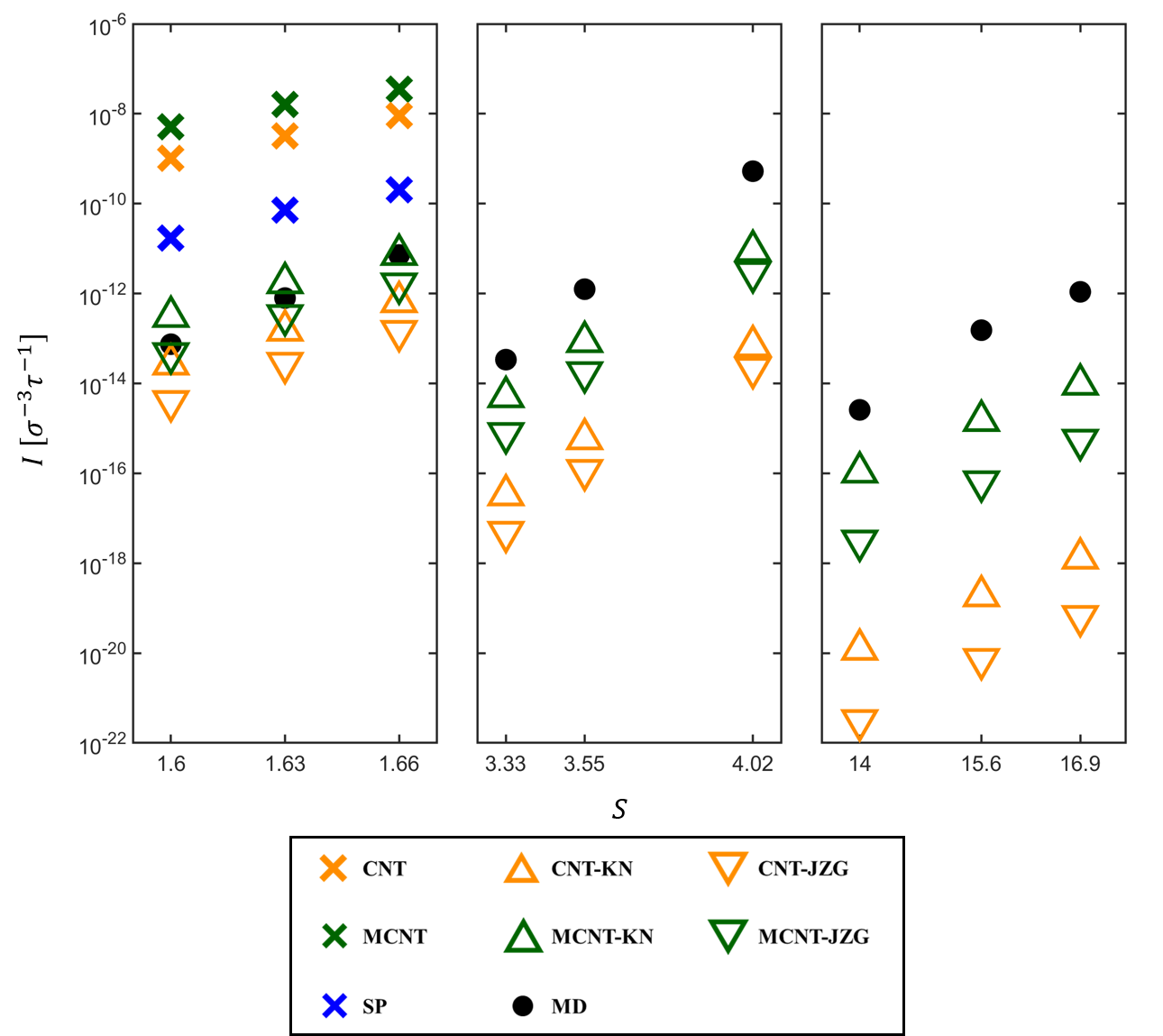}
\caption{Nucleation rates from simulations \cite{diemand_large_2013} and theoretical models. 
The left panel presents nucleation rates at relatively low supersaturation for cases with $T = 1.0\epsilon/k_\mathrm{B}$. Black circles represent MD simulation results \cite{diemand_large_2013}. Results from traditional CNT, MCNT \cite{diemand_large_2013}, and SP \cite{diemand_large_2013} models are shown as orange, dark green, and blue crosses, respectively. Results from our thermodynamic CNT and MCNT models are depicted as orange and dark green triangles, respectively, with up-pointing triangles corresponding to calculations using KN EOS, and down-pointing triangles to those using JZG EOS. 
The middle and right panels illustrate nucleation rates at moderate and high supersaturation, for cases with $T = 0.8\epsilon/k_\mathrm{B}$ and $0.6\epsilon/k_\mathrm{B}$, respectively. Only results from MD simulations \cite{diemand_large_2013} and our thermodynamic models are shown in these two panels.
Although the supersaturation values measured in simulations are not required as input for our thermodynamic models, they are utilized for plotting purposes.}
\label{fig:AllResults}
\end{figure}

The results are summarized in Table \ref{tab:data}. The Nucleation rate is in the unit of $[\sigma^{-3}\tau^{-1}]$ with $\tau = \sigma\sqrt{m/\epsilon}$, and the work of formation is in the unit of $[\epsilon]$.
Nucleation rates with function of supersaturation are illustrated in Figure \ref{fig:AllResults}. Its left, middle, and right panels are for cases at $T = 1.0\epsilon/k_\mathrm{B}$, $0.8\epsilon/k_\mathrm{B}$, and $0.6\epsilon/k_\mathrm{B}$, respectively. 

\begin{sidewaystable}
\caption{Summary of temperature $T$, initial density $\rho_0$, supersaturation $S$, steady-state nucleation rate $I$, critical cluster size $n^*$, and critical work of formation $\Delta\varOmega^*$ for each case. 
Supersaturation $S$ is calculated based on the measurement of total pressure measured from simulations \cite{diemand_large_2013}. $I_\mathrm{MD}$ is measured from simulation results, and $n^*_\mathrm{MD-NT}$ is the critical cluster size derived from $I_\mathrm{MD}$ using the first nucleation theorem \cite{kalikmanov_nucleation_2013},  $n^*_\mathrm{MD-NT} = (\partial \ln{I_\mathrm{MD}} / \partial \ln{S})_T - 1$, as detailed in Ref. \cite{diemand_large_2013}. Subscripts "CNT", "MCNT", and "SP" correspond to results calculated using the traditional CNT, MCNT, and SP models, respectively. Subscripts "- KN" and "- JZG" denote the results obtained from our thermodynamic models by using KN and JZG EOS, respectively.}
\label{tab:data}
\begin{tabular*}{\textheight}{@{\extracolsep\fill}lccccccccc}
\toprule

RunID & T10n62 & T10n60 & T10n58 & T8n30 & T8n25 & T8n23 & T6n80 & T6n73 & T6n65 \\
\midrule

$T\,[\epsilon/k_\mathrm{B}]$ & 1.0 & 1.0 & 1.0 & 0.8 & 0.8 & 0.8 & 0.6 & 0.6 & 0.6 \\
$\rho_0\,[\sigma^{-3}]$ & 0.062 & 0.060 & 0.058 & 0.030 & 0.025 & 0.023 & 0.0080 & 0.0073 & 0.0065 \\
$S$ \cite{diemand_large_2013} & 1.66 & 1.63 & 1.60 & 4.02 & 3.55 & 3.33 & 16.9 & 15.6 & 14.0 \\
 &  &  &  &  &  &  &  &  & \\

$I_\mathrm{MD}\,[\sigma^{-3}\tau^{-1}]$ \cite{diemand_large_2013} & 7.2$\times 10^{-12}$ & 7.9$\times 10^{-13}$ & 7.5$\times 10^{-14}$ & 5.3$\times 10^{-10}$ & 1.2$\times 10^{-12}$ & 3.4$\times 10^{-14}$ & 1.1$\times 10^{-12}$ & 1.5$\times 10^{-13}$ & 2.6$\times 10^{-15}$ \\
$I_\mathrm{CNT}\,[\sigma^{-3}\tau^{-1}]$ & 9.1$\times 10^{-9}$ & 3.3$\times 10^{-9}$ & 1.0$\times 10^{-9}$ & 2.9$\times 10^{-11}$ & 5.9$\times 10^{-13}$ & 5.2$\times 10^{-16}$ & 1.5$\times 10^{-17}$ & 2.3$\times 10^{-18}$ & 1.4$\times 10^{-19}$ \\
$I_\mathrm{CNT-KN}\,[\sigma^{-3}\tau^{-1}]$ & 6.6$\times 10^{-13}$ & 1.5$\times 10^{-13}$ & 2.7$\times 10^{-14}$ & 6.9$\times 10^{-14}$ & 5.8$\times 10^{-16}$ & 3.3$\times 10^{-17}$ & 1.3$\times 10^{-18}$ & 1.9$\times 10^{-19}$ & 1.2$\times 10^{-20}$ \\
$I_\mathrm{CNT-JZG}\,[\sigma^{-3}\tau^{-1}]$ & 1.4$\times 10^{-13}$ & 2.8$\times 10^{-14}$ & 4.0$\times 10^{-15}$ & 2.2$\times 10^{-14}$ & 1.2$\times 10^{-16}$ & 4.9$\times 10^{-18}$ & 6.5$\times 10^{-20}$ & 7.2$\times 10^{-21}$ & 3.2$\times 10^{-22}$ \\
$I_\mathrm{MCNT}\,[\sigma^{-3}\tau^{-1}]$ \cite{diemand_large_2013} & 3.5$\times 10^{-8}$ & 1.6$\times 10^{-8}$ & 5.1$\times 10^{-9}$ & 9.2$\times 10^{-10}$ & 3.4$\times 10^{-11}$ & 2.9$\times 10^{-12}$ & 3.8$\times 10^{-14}$ & 6.1$\times 10^{-15}$ & 4.8$\times 10^{-16}$ \\
$I_\mathrm{MCNT-KN}\,[\sigma^{-3}\tau^{-1}]$ & 7.4$\times 10^{-12}$ & 1.7$\times 10^{-12}$ & 3.1$\times 10^{-13}$ & 9.0$\times 10^{-12}$ & 8.5$\times 10^{-14}$ & 5.0$\times 10^{-15}$ & 9.6$\times 10^{-15}$ & 1.5$\times 10^{-15}$ & 1.1$\times 10^{-16}$ \\
$I_\mathrm{MCNT-JZG}\,[\sigma^{-3}\tau^{-1}]$ & 1.6$\times 10^{-12}$ & 3.2$\times 10^{-13}$ & 4.6$\times 10^{-14}$ & 2.9$\times 10^{-12}$ & 1.7$\times 10^{-14}$ & 7.7$\times 10^{-16}$ & 5.5$\times 10^{-16}$ & 6.5$\times 10^{-17}$ & 3.1$\times 10^{-18}$ \\
$I_\mathrm{SP}\,[\sigma^{-3}\tau^{-1}]$ \cite{diemand_large_2013} & 2.0$\times 10^{-10}$ & 7.3$\times 10^{-11}$ & 1.7$\times 10^{-11}$ & 1.5$\times 10^{-8}$ & 8.5$\times 10^{-10}$ & 1.2$\times 10^{-11}$ & 7.0$\times 10^{-10}$ & 1.9$\times 10^{-10}$ & 3.0$\times 10^{-11}$ \\
 &  &  &  &  &  &  &  &  & \\

$n^*_\mathrm{MD-NT}$ \cite{diemand_large_2013} & 129 & 126 & 108 & 48 & 51 & 49 & 24 & 32 & 38 \\
$n^*_\mathrm{CNT}$ \cite{diemand_large_2013} & 49 & 54 & 60 & 25 & 32 & 38 & 21 & 23 & 25 \\
$n^*_\mathrm{CNT-KN}$ & 116 & 127 & 141 & 39 & 50 & 58 & 23 & 25 & 28 \\
$n^*_\mathrm{CNT-JZG}$ & 128 & 141 & 157 & 42 & 55 & 64 & 27 & 29 & 33 \\
$n^*_\mathrm{SP}$ \cite{diemand_large_2013} & 62 & 68 & 76 & 21 & 29 & 34 & 15 & 16 & 18 \\
 &  &  &  &  &  &  &  &  & \\

$\Delta\varOmega^*_\mathrm{CNT}\,[\epsilon]$ & 12.70 & 13.67 & 14.76 & 13.73 & 16.56 & 18.37 & 17.47 & 18.50 & 20.05 \\
$\Delta\varOmega^*_\mathrm{CNT-KN}\,[\epsilon]$ & 22.24 & 23.63 & 25.31 & 18.58 & 22.10 & 24.27 & 18.94 & 20.00 & 21.50 \\
$\Delta\varOmega^*_\mathrm{CNT-JZG}\,[\epsilon]$ & 23.76 & 25.32 & 27.22 & 19.50 & 23.39 & 25.80 & 20.73 & 21.95 & 23.68 \\
$\Delta\varOmega^*_\mathrm{MCNT}\,[\epsilon]$ & 10.40 & 11.35 & 12.43 & 9.99 & 12.72 & 14.47 & 12.19 & 13.17 & 14.66 \\
$\Delta\varOmega^*_\mathrm{MCNT-KN}\,[\epsilon]$ & 19.82 & 21.20 & 22.86 & 14.68 & 18.11 & 20.24 & 13.60 & 14.61 & 16.05 \\
$\Delta\varOmega^*_\mathrm{MCNT-JZG}\,[\epsilon]$ & 21.32 & 22.88 & 24.76 & 15.58 & 19.38 & 21.74 & 15.32 & 16.48 & 18.17 \\
$\Delta\varOmega^*_\mathrm{SP}\,[\epsilon]$ & 14.32 & 15.46 & 16.75 & 7.48 & 9.82 & 11.34 & 6.22 & 6.91 & 7.96 \\

\botrule
\end{tabular*}
\end{sidewaystable}

At relatively low supersaturation (left panel of Figure \ref{fig:AllResults}), the critical clusters are relatively large ($n>100$), suggesting that the capillary approximation used in theoretical models is reasonable in this regime. 
Traditional CNT and MCNT models significantly overestimate nucleation rates by approximately 4 orders of magnitude. In contrast, our thermodynamic CNT and MCNT models predict nucleation rates aligning very well with MD results. 
This highlights that CNT and MCNT models can provide accurate predictions at low supersaturation when the thermodynamic driving force, $\Delta\mu$, is calculated accurately.
Notably, both our thermodynamic models outperform the SP model, signifying that, as expected, using an appropriate EOS is more effective than relying on the SP model to bypass the ideal gas assumption underlying traditional CNT and MCNT.

Given that our thermodynamic models have been validated at relatively low supersaturation, we focus on comparing their predicted nucleation rates with MD results at moderate and high supersaturation. As illustrated in the middle and right panels of Figure \ref{fig:AllResults}, CNT with both EOS underestimate the nucleation rate. Using the MCNT model leads better agreement between the theoretical predictions and the simulation.
The difference between the predictions of our thermodynamic CNT and MCNT models is observed to increase with the increase of supersaturation, which is expected as the modification in MCNT becomes more significant when the critical cluster size decreases. Both models show larger deviations from MD results at higher supersaturation, reflecting the diminishing validity of the capillary approximation in these conditions. One needs to remind that we use no fitting parameters in this work. As mentioned, the driving force is derived from the EOS and the surface energy is derived from MD calculations \cite{baidakov_metastable_2007}. A potential correction would be to include the dependence of the surface tension with the cluster curvature, i.e. the so-called Tolman effect \cite{Prestipino}. However, we do not make such an attempt in the present analysis for the following reasons. Firstly, there is no consensus on the latter correction, even its sign is debated. Secondly, the clusters observed in MD simulations \cite{diemand_large_2013} at high supersaturation are found to be diffuse and less dense than predicted by the phase diagram, leading to presumably a lower surface energy \cite{Philippe_JCP_2011,Philippe_Phil_mag_2024}. Accounting for the latter effect in CNT (or MCNT) would probably improve the theoretical predictions. This is left for future research.

When comparing the results obtained using different EOS, we observe that KN EOS generally produce nucleation rates that are closer to the MD results than those obtained using JZG EOS. This finding is consistent with the conclusion drawn in Ref. \cite{stephan_review_2020}, which states that KN EOS are generally more precise than JZG EOS. 
While the difference in chemical potential between these two series of EOS is relatively minor, it can still lead to significant variations in the calculated nucleation rates, sometimes differing by as much as 2 orders of magnitude. This variation arises from the high sensitivity of nucleation rates to the critical work of formation. Additionally, it is crucial to recognize that in the nucleation rate formula, the critical work of formation is divided by $k_\mathrm{B}T$, meaning that lower temperatures will further amplify this sensitivity.

\section{Conclusion}\label{sec:Conclusion}

In this work, we have successfully integrated the calculation of driving force for nucleation using EOS into the framework of CNT and MCNT for determining nucleation rates. We applied this method to the Lennard-Jones system and compared our predictions with MD simulation results. Our approach demonstrates excellent agreement with MD data at relatively low supersaturation, where the capillary approximation remains reasonable, and continues to perform relatively well even at moderate and high supersaturation. This performance surpasses traditional models that rely on the logarithm of supersaturation for calculating the driving force for nucleation. 
Additionally, our findings reveal that using more precise EOS generally results in better alignment with MD data. Looking ahead, we plan to extend our comparison to MD simulations employing the seeding technique at lower supersaturation, focusing on not only nucleation rates but also other nucleation properties such as cluster size and cluster density.

\bmhead{Acknowledgments}
The authors thank Aymane Graini and Yilin Ye for fruitful discussion.
This work was supported by the ANR TITANS project.

\bibliography{sn-bibliography}% common bib file

\end{document}